\documentclass[epsfig,graphics,twocolumn,showpacs,floatfix,mathbbm,prl,superscriptaddress]{revtex4}
\usepackage{amsmath,amssymb,graphicx,color}
\usepackage{float}
\usepackage{subfigure}
\newcommand{\PrYSO}{Pr$^{3+}$:Y$_2$Si{O$_5$}}
\newcommand{\YSO}{Y$_2$Si{O$_5$}}

\graphicspath{{figs/}{}}

\begin{document}

\title{A spectral hole memory for light at the single photon level}

\author{Kutlu Kutluer}
\affiliation{ICFO-Institut de Ciencies Fotoniques, The Barcelona Institute of Technology, Mediterranean Technology Park, 08860 Castelldefels (Barcelona), Spain}
\author{Mar\'{i}a Florencia Pascual-Winter}
\affiliation{Laboratorio de Fot\'{o}nica y Optoelectr\'{o}nica, Centro At\'{o}mico Bariloche and Instituto Balseiro, C.N.E.A., (8400) San Carlos de Bariloche, Argentina}
\author{Julian Dajczgewand}
\affiliation{Laboratoire Aim\'{e} Cotton UMR 9188, CNRS, Universit\'{e} Paris-Sud and ENS Cachan, B\^{a}timent 505, Campus Universitaire, 91405 Orsay, France}
\author{Patrick M. Ledingham}
\altaffiliation{Present address: Department of Physics, Clarendon Laboratory, University of Oxford, Oxford OX1 3PU, United Kingdom}
\author{Margherita Mazzera}
\email{margherita.mazzera@icfo.es}
\affiliation{ICFO-Institut de Ciencies Fotoniques, The Barcelona Institute of Technology, Mediterranean Technology Park, 08860 Castelldefels (Barcelona), Spain}
\author{Thierry Chaneli\`{e}re}
\affiliation{Laboratoire Aim\'{e} Cotton UMR 9188, CNRS, Universit\'{e} Paris-Sud and ENS Cachan, B\^{a}timent 505, Campus Universitaire, 91405 Orsay, France}
\author{Hugues de Riedmatten}
\affiliation{ICFO-Institut de Ciencies Fotoniques, The Barcelona Institute of Technology, Mediterranean Technology Park, 08860 Castelldefels (Barcelona), Spain}
\affiliation{ICREA-Instituci\'{o} Catalana de Recerca i Estudis Avan\c cats, 08015 Barcelona, Spain}

\begin{abstract}
We demonstrate a solid state spin-wave optical memory based on stopped light in a spectral hole. A long lived narrow spectral hole is created by optical pumping in the inhomogeneous absorption profile of a \PrYSO \, crystal. Optical pulses sent through the spectral hole experience a strong reduction of their group velocity and are spatially compressed in the crystal. A short Raman pulse transfers the optical excitation to the spin state before the light pulse exits the crystal, effectively stopping the light. After a controllable delay, a second Raman pulse is sent, which leads to the emission of the stored photons. We reach storage and retrieval efficiencies for bright pulses of up to $39\,\%$ in a $5 \,\mathrm{mm}$-long crystal. We also show that our device works at the single photon level by storing and retrieving $3\,\mathrm{\mu s}$-long weak coherent pulses with efficiencies up to $31\,\%$, demonstrating the most efficient spin-wave solid state optical memory at the single-photon level so far. We reach an unconditional noise level of $(9\pm1)\times 10^{-3}$ photons per pulse in a detection window of $4\,\mathrm{\mu s}$ leading to  a signal-to-noise ratio of $33 \pm 4$ for an average input photon number of 1, making our device promising for long-lived storage of non-classical light.
\end{abstract}

\maketitle
Coherent interactions between light and atoms play an important role in modern quantum science \cite{Hammerer2010a}. They enable the control of the properties of light, e.g. its velocity, until the extreme case where the light is stopped. This allows the realization of quantum memories for light, which are fundamental building blocks of complex quantum information processing protocols \cite{Bussieres2013,Duan2001,Sangouard2011,Kimble2008,Knill2001}. Their implementation requires the mapping of a single photon light field onto long-lived atomic coherence.

Rare-earth doped solids are attractive for quantum memory applications because they contain a large number of naturally trapped atoms with long spin and optical coherence times. Several experiments have demonstrated the storage of quantum information carried by weak coherent light states in the optically excited state of rare-earth ions using the atomic frequency comb (AFC)  \cite{Riedmatten2008, Usmani2010, Sabooni2010,Gundogan2012,Zhou2012, Sinclair2014} or gradient echo memory (GEM) technique \cite{Lauritzen2010,Hedges2010}. Non-classical states of light have also been stored in excited states using AFC \cite{Clausen2011,Saglamyurek2011,Clausen2012, Rielander2014,Saglamyurek2015}. However, this  leads to short and mostly predetermined storage times. Longer storage times and on-demand read-out can be achieved by transfering the optical collective excitation into spin collective excitations (spin waves) using strong Raman pulses \cite{Afzelius2010, Gundogan2013}. AFC spin-wave storage has been extended to the quantum regime recently \cite{Gundogan2015,Jobez2015}, with storage and retrieval efficiencies of a few percents.

Another well-known technique for quantum storage is based on Electromagnetically-Induced Transparency (EIT) \cite{Chaneliere2005,Eisaman2005,Choi2008,Zhou2012a}. It relies on the creation of a narrow transparency window in an otherwise opaque medium, thanks to the application of a control field. Stopped light based on EIT has also been demonstrated in rare-earth doped crystals (\PrYSO) \cite{Turukhin2001}, leading (together with dynamical decoupling techniques) to ultra-long storage time for bright pulses \cite{Longdell2005}, up to the regime of one minute \cite{Heinze2013}, the longest light  storage time demonstrated so far. However, the efficiency for these experiments was also quite low (few percent at most). Only recently a multipass configuration has been implemented in \PrYSO\, to increase the effective optical depth of the medium and to store classical pulses for a few $\mathrm{\mu s}$ with an internal efficiency of about $76\,\%$ \cite{Schraft2016}. Nonetheless, EIT has not yet been demonstrated at the single photon level in doped crystals. This may prove challenging, because of the noise induced by the control field during the read-out.

In this paper, we demonstrate a simple and robust alternative protocol to store and retrieve light in the spin-state of a doped crystal, that enables us to reach high efficiencies and single photon level operation. The technique, proposed by Lauro et al. \cite{Lauro2009}, uses a permanent transparency window created in a doped crystal by spectral hole burning. While the protocol is similar to EIT because it is based on slow light, there are two important differences. First, the transparency window is not created dynamically with a control pulse, but by optical pumping way before the photons to be stored enters the medium. Second, the photons excite optical coherence with off resonance atoms. There is therefore no dark-state as in EIT. Slow light experiments based on spectral holes have been performed before with bright pulses \cite{Sellars2003, Hahn2008, Ham2009, Lauro2009a}, but without the possibility to store. In this paper, the storage mechanism is based on the sequential conversion of the optical coherence into a spin coherence, using short Raman $\pi$ pulses. This is important in practice because the Raman pulses can be much shorter than the retrieved single photon, which enables temporal filtering. This greatly facilitates the operation at the single-photon level. Thanks to the robust memory preparation, we demonstrate storage and retrieval efficiencies up to $39\,\%$ for bright pulses. In addition, we store and retrieve weak coherent pulses at the sub-photon level with an efficiency of $31\,\%$, the highest achieved so far for a single-photon level solid state spin-wave optical memory \cite{Jobez2015,Gundogan2015}. We reach an unconditional noise floor of $(2.25\pm0.25)\times 10^{-3}$\, photons/$\mathrm{\mu s}$. For a detection window of $4 \, \mathrm{\mu s}$, this leads to a signal-to-noise ratio of  $33 \pm 4$ for an average input photon number of 1 (together with a slightly reduced efficiency of $23\,\%$), the highest demonstrated so far in a crystal. 

\begin{figure}[ht]
\centering
\subfigure[]{
\centering\includegraphics[width=0.45\textwidth]{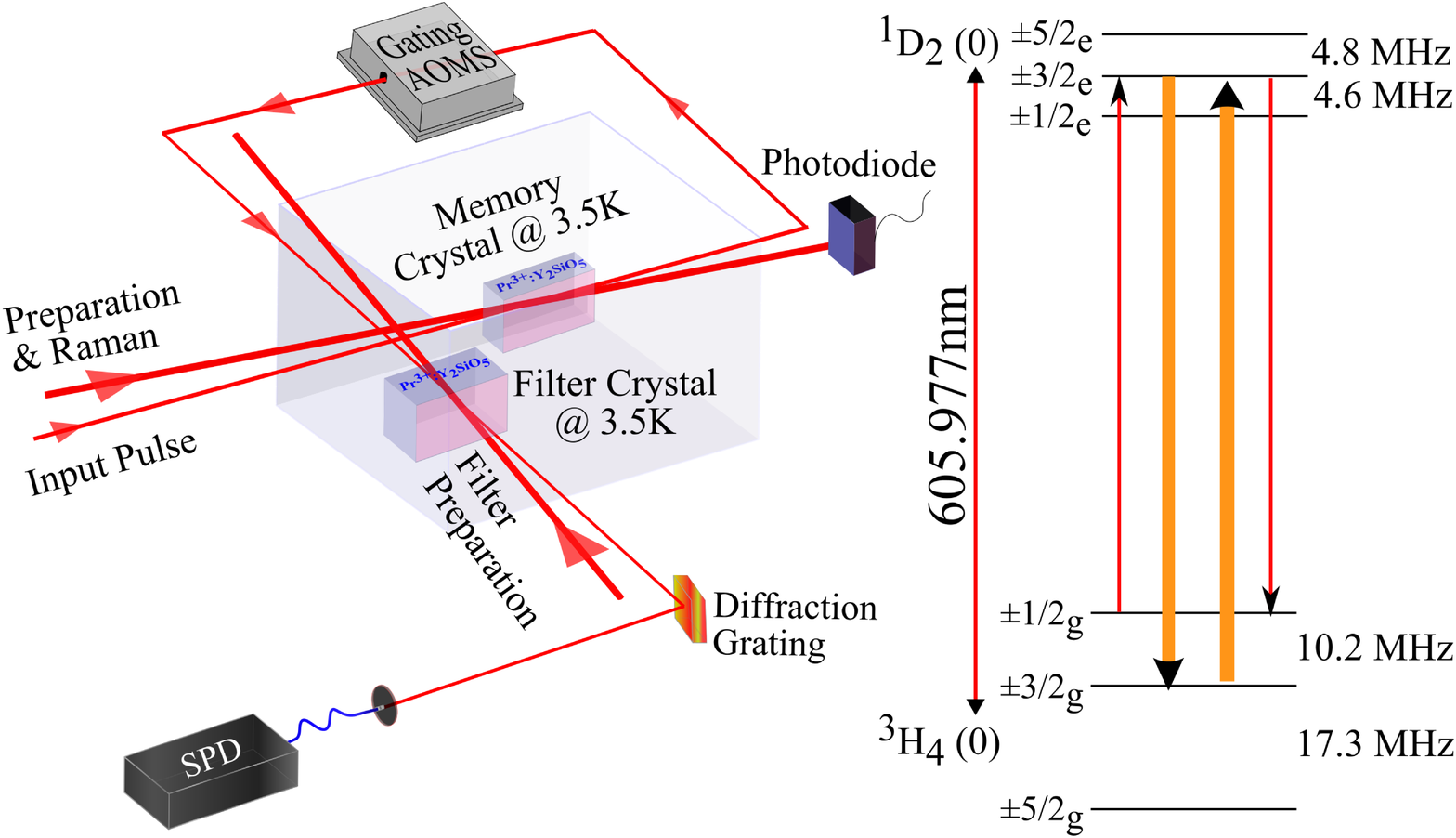}
    \label{fig:subfig1}
}
\subfigure[]{
\centering\includegraphics[width=0.45\textwidth]{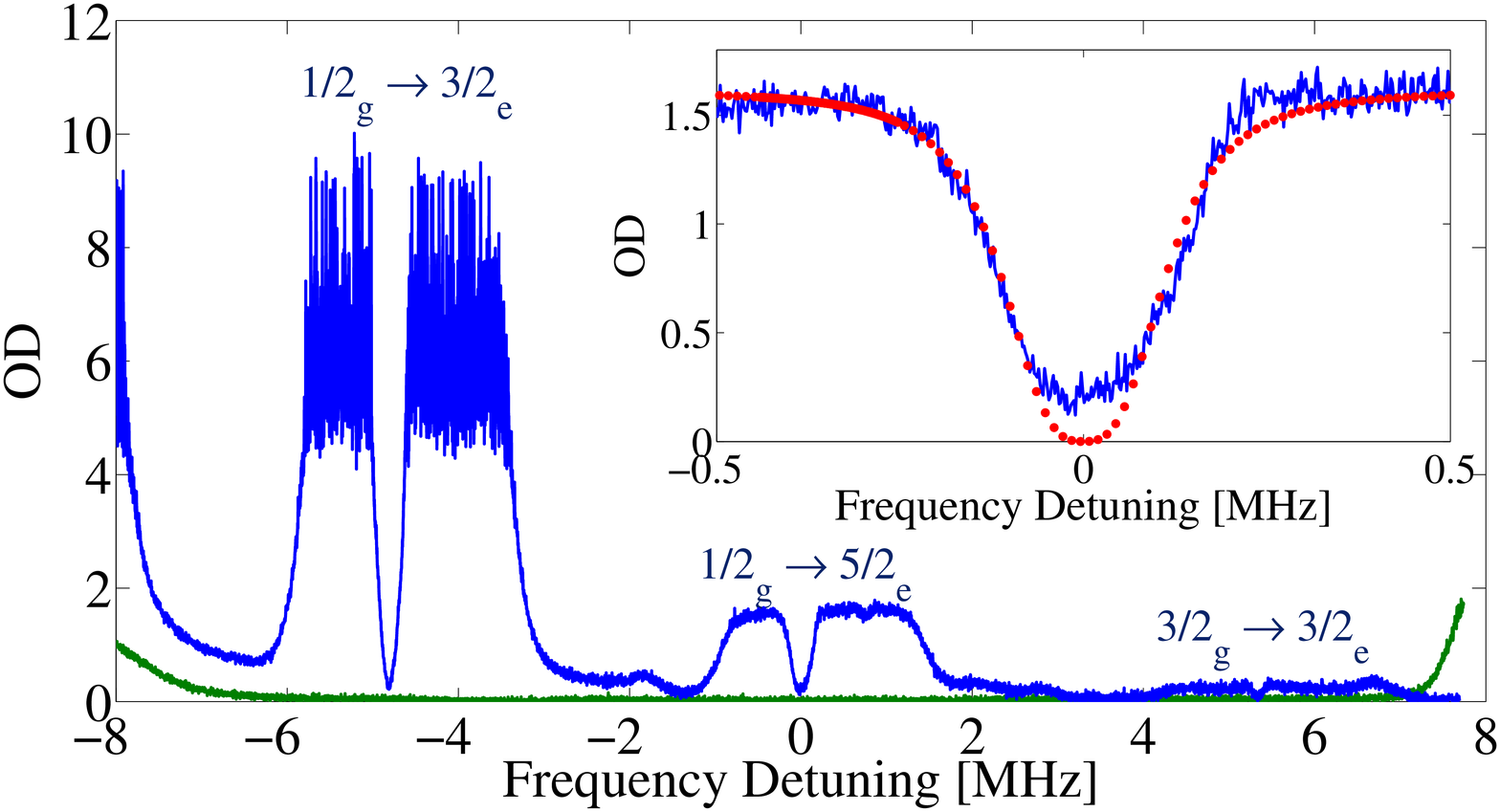}
    \label{fig:subfig2}
}

\caption{(a) Left Panel: experimental setup. The preparation/Raman and input beams arrive at the memory with an angle of $\sim 4\,\mathrm{^{\circ}}$. The polarization is rotated along the D$_{2}$ crystal axis in order to maximize the absorption.
 Right Panel: hyperfine splitting of the first sublevels (0) of the ground $^{3}$H$_{4}$ and the excited $^{1}$D$_{2}$ manifold of Pr$^{3+}$ in \YSO.
(b) Spectral holes burnt in $2.1\,\mathrm{MHz}$-wide single class absorption features. Inset: magnification of the spectral hole resonant with the ${1}/{2}_\textrm{g}-{5}/{2}_\textrm{e}$ transition. The fit of the hole shape  (see text for the details) is also shown. }
\end{figure}

The experimental setup is shown in Fig. \ref{fig:subfig1}. The source of coherent light at $606 \,\mathrm{nm}$ is a Toptica TA-SHG-pro laser, frequency stabilized by means of the Pound-Drevel-Hall technique with a home-made Fabry-Perot cavity in vacuum. Before the memory, it is split into two beams, one used as input beam ($150\,\mathrm{\mu W}$ of power and diameter of $70\,\mathrm{\mu m}$, at the memory crystal) and the other ($20\,\mathrm{mW}$ of power and diameter of $300\,\mathrm{\mu m}$, at the memory crystal) employed for the memory preparation and for the Raman pulses. Both are frequency- and amplitude-modulated with double pass acousto-optic modulators, driven by a Signadyne arbitrary waveform generator. The two beams overlap on the memory crystal, a $0.05 \,\%$ \PrYSO \, sample of length $L = 5 \,\mathrm{mm}$, located in a closed cycle cryostat operating at $3.5\, \mathrm{K}$ (Cryostation, Montana Instruments). The optical transition of interest for the storage of light, the $^{3}H_{4}(0) \rightarrow ^{1}D_{2}(0)$, is characterized by an inhomogeneous width of $6\,\mathrm{GHz}$. After the storage, the retrieved signal is coupled in a polarization maintaining single-mode fiber and collected with a photodetector.

The memory is preparated by burning a narrow spectral hole in a  $2.1\,\mathrm{MHz}$  wide absorption feature at the frequency of the ${1}/{2}_\textrm{g} \rightarrow {3}/{2}_\textrm{e}$ transition, created using the procedure described in \cite{Maring2014, Gundogan2015}. Compared to a bare hole burning procedure  \cite{Lauro2009a}, this technique has three advantages. It permits to address only a single class of ions (thus allowing the Raman pulses to address only one transition), to empty the spin storage state (${3}/{2}_\textrm{g}$), and also to control the optical depth of the spectral hole, by varying the burn back power used to create the single class absorption feature. An example of absorption trace with a spectral hole, about $\Delta_0$ = $230 \,\mathrm{kHz}$ wide and of optical depth $\alpha L = 8.7$,  is provided in Fig. \ref{fig:subfig2}. Due to a limited dynamical range of the photodetector, we cannot directly measure the optical depth of the spectral hole at the ${1}/{2}_\textrm{g} \rightarrow {3}/{2}_\textrm{e}$ transition. We extrapolate the value by fitting the hole on the ${1}/{2}_\textrm{g} \rightarrow {5}/{2}_\textrm{e}$ transition (shown in the inset of Fig. \ref{fig:subfig2}), which is not affected by the detector non-linearity, and applying a scaling factor according to the relative oscillator strength of the two optical transitions \cite{Nilsson2004}. The validity of our approach is tested by preparing weakly absorbing features and comparing the directly measured optical depth with the one extrapolated.

\begin{figure}[h]
\centerline{\includegraphics[width=1\columnwidth]{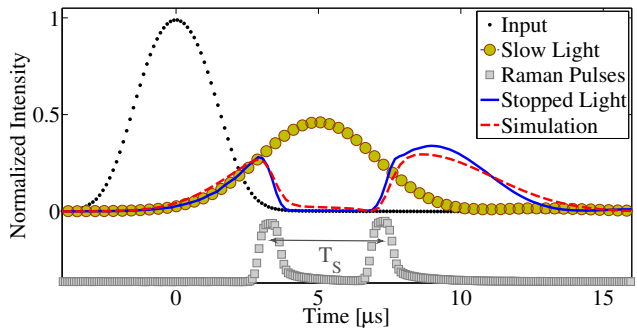}}
\caption{The spectral hole memory scheme. Black dotted trace: input pulse; green solid circles: slowed pulse; blue solid trace: stopped light; grey square trace: Raman pulses (arbitrary units); red dashed trace: the numerical simulation of stopped light.}
\label{echoes}
\end{figure}

The light storage sequence is depicted in Fig. \ref{echoes}. An input pulse is sent resonantly to the spectral hole prepared in the ${1}/{2}_\textrm{g} \rightarrow {3}/{2}_\textrm{e}$ transition. The Gaussian pulse has a duration of $ 3\,\mathrm{\mu s}$ (full width at half maximum, FWHM). The black dot trace in Fig. \ref{echoes} represents the input pulse, linearly polarized perpendicular to the $D_2$ axis to minimize the absorption, traveling through a $18 \,\mathrm{MHz}$-wide transparency window. We assume it undelayed and take it as a reference. When the pulse penetrates through the spectral hole presented in  Fig. \ref{fig:subfig2}, it is slowed down by approximately $5\, \mathrm{\mu s}$ (green solid circles in Fig. \ref{echoes}). The delayed pulse is stretched and slightly attenuated with respect to the  input pulse because its bandwidth $\sim 150\,\mathrm{kHz}$ FWHM  is comparable to the more squarish hole width $\Delta_0=230 \,\mathrm{kHz}$. A longer pulse would be less stretched but not sufficiently separated from the input for the spin storage step \cite{Lauro2009}. 

The shape of the hole is mostly due to  frequency jitter of the laser. To account for it, we fit the absorption profile to a $hyperlorentzian$ function given by the equation
\begin{equation}
g\left(\Delta\right)=1- \frac{1}{1+\left|2\Delta/\Delta_0\right|^n}
\label{g_fit}
\end{equation} ($n=2$ for a lorentzian), where $\Delta$ is the frequency detuning. From the fit, we obtain the previously mentioned values $\Delta_0 = 230\, \mathrm{kHz}$ with $n=3.0$ and $\alpha L = 8.7$ (see the red dashed trace in the inset of Fig. \ref{fig:subfig2}).

We then transfer the optical collective excitation to a spin-wave using a short Raman pulse (grey square trace in Fig. \ref{echoes}) on the ${3}/{2}_\textrm{e} \rightarrow {3}/{2}_\textrm{g}$ transition. After a controllable time $T_s$, a second Raman pulse triggers the pulse emission by reconverting the spin-wave into an optical excitation that will slowly propagate through the crystal. This is shown with the blue solid trace in Fig. \ref{echoes}. Since the delay of the slow light is not sufficient to completely compress the initial pulse into the crystal, some light escapes before we send the first Raman pulse. For this measurement, the storage and retrieval efficiency $\eta_S$, calculated as the ratio between the areas of the retrieved pulse (after the second Raman pulse) and the input pulse, is $39\,\%$.

The pulse propagation and storage is modeled by the Schr\"odinger-Maxwell equations in one dimension (along $z$). For three-level atoms, the rotating-wave probability amplitudes $C_g$, $C_e$ and $C_s$  for the ground (${1}/{2}_\textrm{g}$), excited (${3}/{2}_\textrm{e}$) and spin (${3}/{2}_\textrm{g}$) states, respectively, are governed by the time-dependent Schr\"odinger equation:
\begin{eqnarray}
i \partial_t \left[
\begin{array}{c}
C_g \\
C_e \\
C_s\\
\end{array}\right]
=
\left[
\begin{array}{ccc}
0 & \mathcal{E}/2 & 0 \\
\mathcal{E}/2 & -\Delta & \Omega/2 \\
0 & \Omega/2 &0 \\
\end{array}\right]
\left[
\begin{array}{c}
C_g \\
C_e \\
C_s\\
\end{array}\right]
\label{bloch}\end{eqnarray}
where $\mathcal{E}(z,t)$  is  the envelope of the input signal. As a consequence, $C_g$, $C_e$ and $C_s$ depend on $z$ and $t$ for a given detuning $\Delta$ within the  inhomogeneous broadening. Raman pulses applied on the ${3}/{2}_\textrm{e} \rightarrow{3}/{2}_\textrm{g}$ are described by the envelope $ \Omega(t)$ which does not depend on $z$ because the state ${3}/{2}_\textrm{g}$ is initially empty (no absorption). The Raman beam satisfies the two-photon resonance condition. We  neglect the decoherence in this first approach since we are mostly interested in modeling the efficiency for storage times shorter than the coherence time.

The propagation of the signal $\mathcal{E}(z,t)$ is described by the Maxwell equation that can be simplified in the slowly varying envelope approximation:
\begin{equation}
\partial_z\mathcal{E}(z,t)+ \frac{1}{c}\partial_t\mathcal{E}(z,t)=
-\displaystyle\frac{i \alpha}{2\pi}\int_\Delta g\left(\Delta\right) C_g C_e^* \mathrm{d}\Delta \label{MB_M}
\end{equation}
The term $C_g C_e^* $ is the atomic coherence on the ${1}/{2}_\textrm{g} \rightarrow {3}/{2}_\textrm{e}$ transition, proportional to the atomic polarization. The light coupling constant is directly included in the absorption coefficient $\alpha$.
The Schr\"odinger-Maxwell equations (\ref{bloch},\ref{MB_M}) can be further simplified because the signal is weak. In the perturbative regime $C_g \simeq 1$, so the atomic evolution is only described by $C_e$ and $C_s$.

Slow-light propagation (without Raman pulses) can be described analytically because in that case $C_s=0$ thus reducing eq. \eqref{bloch} to the evolution of $C_e$ only \cite{Lauro2009a}. For the storage step (with Raman pulses), the Schr\"odinger-Maxwell equations are integrated numerically.
 For a given inhomogeneous detuning  $\Delta$, we calculate the atomic evolution \eqref{bloch}  by using a fourth-order Runge-Kutta method. After integrating over the inhomogeneous broadening using the $hyperlorentzian$ function \eqref{g_fit} for $g\left(\Delta\right)$, the propagation equation \eqref{MB_M} is integrated along $z$ using the Euler method. The stopped light temporal profile can be well reproduced by our numerical simulation (red dashed trace) in Fig. \ref{echoes}. To account for a possible imperfection of the Raman transfer to the spin state, we have adjusted the Raman pulse area to $0.85\pi$ instead of $\pi$. This artificially incorporates the decoherence mechanism that is not included in our model.

\begin{figure}
\centering\includegraphics[width=1\columnwidth]{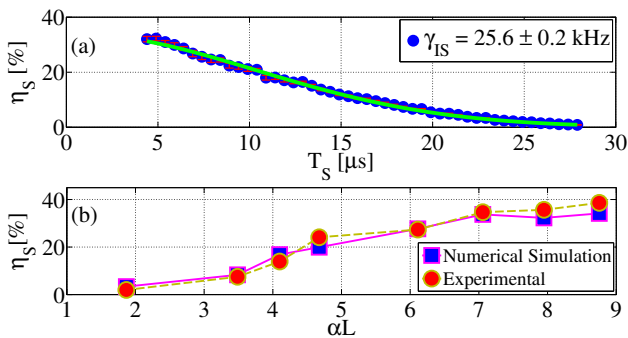}
\caption{ (a) Storage efficiencies as a function of the delay $T_s$ between the two Raman pulses. The decay is fitted with a Gaussian profile and the inhomogeneous spin broadening of $\gamma_{IS} = (25.6 \pm 0.2) \,\mathrm{kHz}$ is extracted. (b) Storage efficiencies (experimental: red circles; numerical simulation: blue squares) as a function of the hole optical depth.}
\label{eff}
\end{figure}

In order to characterize the storage, we investigate the efficiency of the stopped light as a function of the delay between the two Raman pulses, $T_{s}$, and of the hole optical depth. We show in Fig. \ref{eff}(a) that the decay of the signal is compatible with the inhomogeneous broadening of the spin transition (with inhomogeneous linewidth $\gamma_{IS} = (25.6 \pm 0.2) \,\mathrm{kHz}$), confirming that the pulse energy is stored as a spin wave \cite{Afzelius2010, Gundogan2013, Gundogan2015}. In Fig. \ref{eff}(b) we compare the experimental values of storage efficiencies (red circles) and the results of the numerical simulations (blue squares) as functions of the optical depth $\alpha L$. For these measurements, the position of the first Raman pulse is always optimized in order to maximize the efficiency since the group delay decreases at lower optical depths. We observe that the efficiency grows steadily as a function of optical depth. The experimental measurements are well reproduced by the numerical simulations, thus supporting our analysis.

In order to test the suitability of our optical memory to work in the quantum regime, we insert neutral density filters in the input mode to decrease the intensity of the input pulses down to the single-photon level. For these measurements, we perform 1000 storage and retrieval trials for each memory preparation, at a rate of $5 \,\mathrm{kHz}$. In order to discriminate the retrieved single-photon-level signals from the noise originated from the Raman pulses, we apply several filtering strategies (see Fig. \ref{fig:subfig1}) \cite{Gundogan2015}. First of all, the input and the preparation/Raman modes are spatially separated with an angle of about $4\,\mathrm{^{\circ}}$. After the memory, the retrieved signal is first steered to two single pass AOMs acting as temporal gates. Then it is sent to a second \PrYSO \, crystal where a $1\,\mathrm{MHz}$ transparency window is created at the input pulse frequency. 

\begin{figure}
\centering\includegraphics[width=1\columnwidth]{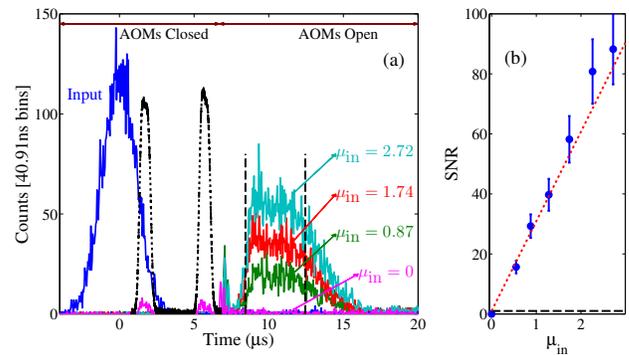}
\caption{(a) Time histograms of the retrieved photons for different input photon numbers $\mu_{in}$  for a spectral hole with $\Delta_0= 275\,\mathrm{kHz}$  in the memory crystal. The input with $\mu_{in} = 0.87$ (blue solid trace) and the Raman pulses (black dotted trace), as measured in photon counting and from a reference photodiode, respectively, are also displayed. The detection window $\Delta t_d = 4\,\mathrm{\mu s}$ is indicated by the dashed lines about the retrieved signal. (b) SNR as a function of  $\mu_{in}$. The error bars are evaluated with Poissonian statistics. The black dashed line indicates the limit of detection $SNR = 1$. The dotted line is a linear fit of the data, from which we obtain $\mu_1$ value of $0.030 \pm 0.004$ can be obtained.}
\label{SNR}
\end{figure}

The fluorescence not resonant with the Pr$^{3+}$  $^{3}$H$_{4}(0)  \rightarrow ^{1}$D$_{2}(0)$ transition is then suppressed with a diffraction grating. Finally, the stored and retrieved light is detected with a single-photon detector (SPD). The total transmission of the input light from the cryostat window until the SPD is around $15\,\%$. We record the arrival times of the photons and reconstruct the time histogram for different input photon numbers $\mu_{in}$, as shown in Fig. \ref{SNR}. From the trace  with $\mu_{in}=0$, we measure an unconditional noise floor of $(9\pm1)\times 10^{-3}$ photons per pulse in a detection window $\Delta t_d = 4\,\mathrm{\mu s}$. $\eta_S$ is around $23\,\%$ for $\Delta t_d = 4\,\mathrm{\mu s}$ and 31$\%$ for $\Delta t_d = 7\,\mathrm{\mu s}$. These are the highest efficiencies obtained so far for solid state spin-wave optical memories at the single-photon level. Fig. \ref{SNR}(b) shows the behavior of the signal-to-noise ratio ($SNR$) of the retrieved photons as a function of $\mu_{in}$. We measure a $SNR$ of $33 \pm 4$  ($23 \pm 3$) for $\mu_{in} = 1$ when $\Delta t_d$ = 4$\,\mathrm{\mu s}$ ($\Delta t_d$ = 7$\,\mathrm{\mu s}$), the highest values measured so far for a single photon level solid state spin-wave memory \cite{Gundogan2015, Jobez2015}.

An important figure of merit for a single-photon-level optical memory is the parameter $\mu_1=\mu_{in}/SNR$, which gives the minimum $\mu_{in}$ to reach a $SNR=1$ for the retrieved photon. It has been shown that, in order to achieve quantum storage with an external single-photon source, a necessary condition is to have $p>\mu_1$, where $p$ is the probability to find a single photon before the memory \cite{Gundogan2015,Jobez2015}. From the slope in Fig. \ref{SNR} (b), we find $\mu_1$=$0.030 \pm 0.004$. Our system is therefore very promising for quantum light storage, provided that $\mu s$-long single photons are available. Such long photons could be created from atomic ensembles \cite{Zhao2014, Farrera2016} or single ions \cite{Stute2012} and frequency shifted to the resonance frequency of the Pr$^{3+}$ doped crystal by quantum frequency conversion techniques \cite{Radnaev2010,Albrecht2014,Maring2014}. Shorter photons could also be stored if larger holes are prepared. However, in order to keep a sufficient separation between the second Raman pulse and the emitted photon, shorter Raman pulses should be used, which in turn will require a larger Rabi frequency. This could be achieved by increasing the Raman pulse power, or more efficiently by confining the interaction, e.g. in a waveguide configuration \cite{Corrielli2016}.

We note that the efficiency obtained in this work was mostly limited by the available optical depth and by the limited transfer efficiency. Higher optical depths will lead to higher efficiencies. Numerical simulations show that with a two-times larger optical depth $\alpha L=17.5$, we would reach an efficiency of $55\,\%$ (assuming $100\,\%$ transfer efficiency), for a properly adjusted input pulse duration (1.6 times longer than for the current paper). This shows that our protocol has a favourable scaling for an increasing optical depth. It should be also noted that further improvements could be reached by optimizing the temporal shape of the input and control pulses and the spectral shape of the spectral hole. Further modelling is required.

In our experiment, we store and retrieve single mode weak pulses. When extended to the storage of true single photons, the protocol could be exploited to demonstrate entanglement between remote crystals using the scheme proposed in \cite{Simon2007}. Nonetheless, many applications require the storage and retrieval of photonics qubits. While our protocol is not a good candidate to store light in multiple temporal modes \cite{Nunn2008}, it could be readily extended to the storage of polarization qubits \cite{Gundogan2012,Clausen2012,Zhou2012,Laplane2015} or to frequency-bin qubits \cite{Olislager2010,Sinclair2014}.

To conclude, we implemented a new light storage protocol based on stopped light in a spectral hole in a doped crystal and we achieved a storage and retrieval efficiency of up to $39\, \%$. Thanks to a low unconditional noise floor, we store and retrieve single-photon-level pulses with high signal-to-noise ratio. This demonstrates that the memory can work in the quantum regime, with the highest efficiency so far obtained for spin-wave solid state optical  memories. These results are promising for the realization of robust, highly efficient and long-lived spin-wave solid state quantum memories.

\textbf{Acknowledgments.} This work was supported by the People Programme (Marie Curie Actions) of the EU FP7 under REA
Grant Agreement No. 287252.

Research at ICFO is supported by the ERC starting grant QuLIMA, by the Spanish
Ministry of Economy and Competitiveness (MINECO) and the Fondo
Europeo de Desarrollo Regional (FEDER) through grant
FIS2012-37569, by MINECO Severo Ochoa through grant SEV-2015-0522, by AGAUR via 2014 SGR 1554, and by Fundaci\'o Privada Cellex.

Research at Laboratoire Aim\'{e} Cotton is supported by the French national grants ANR-12-BS08-0015-02 (RAMACO) and ANR-14-CE26-0037-02 (DISCRYS).

\bibliographystyle{prsty}

\end{document}